\documentclass[showpacs,aps,prb,twocolumn,superscriptaddress]{revtex4}
\usepackage{amsmath}
\usepackage{graphicx} 
\usepackage{dcolumn}
\usepackage{bm}
\usepackage{amssymb,amsmath}

\usepackage{multirow}
\usepackage{natbib}

\begin{document}

\title{Anomalous magneto-optical Kerr hysteresis loops in Fe$_{0.25}$TaS$_{2}$}

\author{Chanjuan Sun}
\author{Junichiro Kono}
\affiliation{Department of Electrical and Computer Engineering, Rice University, Houston, Texas 77005, USA}
\affiliation{Department of Physics and Astronomy, Rice University, Houston, Texas 77005, USA}

\author{Adilet Imambekov}
\author{Emilia Morosan}
\affiliation{Department of Physics and Astronomy, Rice University, Houston, Texas 77005, USA}

\date{\today}

\begin{abstract}
We have performed magneto-optical Kerr spectroscopy measurements on intercalated transition-metal dichalcogenide Fe$_{0.25}$TaS$_{2}$ in the polar Kerr geometry as a function of temperature, magnetic field, and wavelength.  The Kerr angle exhibits pronounced peaks at $\sim$775~nm (1.6~eV) and $\sim$515~nm (2.4~eV), which we attribute to spin-dependent interband optical transitions arising from states in the vicinity of the Fermi energy.  Below the ferromagnetic transition temperature (165~K), we observe strongly wavelength- and magnetic-field-dependent Kerr signal.  At a fixed wavelength, the magnetic-field dependence of the Kerr angle shows a clear hysteresis loop, but its shape sensitively changes with the wavelength.  We propose a model that takes into account contributions from domain walls, which allowed us to derive a mathematical expression that successfully fits all the observed hysteresis loops.
\end{abstract}


\maketitle


\section{Introduction}\label{sec:Intro}
There has been much interest in the unique electronic and magnetic properties of transition-metal dichalcogenide (TMDC) compounds, TX$_{2}$, where T is a transition-metal atom and X is sulfur, selenium, or tellurium.\cite{WilsonYoffeAP1969,IntercMaterLevy}  The extremely anisotropic crystal structures of these layered materials provide a natural quasi-two-dimensional platform for the fundamental study of electronic transport and magnetism in low dimensions.  In addition, the weak van der Waals interlayer bonding allows for intercalation of a variety of atoms, ions, and molecules,\cite{NaritaJSSCh1994,EibschutzJAP1981,ParkinPB1980,GuoJPC1987,CyrusPRB1982,MotizukiJMMM1992} which creates an unusually rich family of materials, encompassing insulators, semiconductors, semimetals, normal metals, and superconductors.  Charge density wave (CDW) states sometimes coexist, and compete with, superconductivity in a number of these materials,\cite{VallaPRL2004} while long-range magnetic order can occur when TX$_{2}$ is intercalated with 3$d$-transition metals such as Mn and Fe.

In Fe-intercalated TaS$_{2}$, the Fe atoms form a superlattice when the atomic ratio $x$ of Fe is either 1/3 or 1/4.\cite{NaritaJSSCh1994, MorosanPRB2007Fe}  X-ray diffraction measurements of Fe$_{0.25}$TaS$_{2}$ show that TaS$_2$ has 2H structure and Fe ions occupy the octahedral sites between TaS$_2$ layers, with a hexagonal axis length
$a' = 2a_0$,\cite{NaritaJSSCh1994, MorosanPRB2007Fe} where $a_0$ is the basic hexagonal lattice parameter of TaS$_2$.  Ferromagnetic order occurs in Fe$_{0.25}$TaS$_{2}$ below 160~K.~\cite{EibschutzJAP1981, MorosanPRB2007Fe, VannettePRB2009}  Extremely sharp switching behavior in magnetization versus magnetic field curves has been observed, and coercive fields as high as 3.7~T have been measured at 2~K.\cite{MorosanPRB2007Fe}  This sharp reversal of magnetization at low temperature was utilized to determine the ordinary and anomalous Hall coefficients in Hall measurements.\cite{CheckelskyPRB2008}  More recently, the magnetic domain structure of single crystal Fe$_{0.25}$TaS$_{2}$ was studied by magneto-optical (MO) Faraday effect.\cite{VannettePRB2009}  Real-time MO images revealed unusual dendritic domain structures and slow dynamics of domain formation and propagation.  However, to fully elucidate spin-dependent electronic states in these compounds, wavelength-dependent MO measurements are needed.

In this paper, we report results of detailed magneto-optical Kerr effect (MOKE) spectroscopy measurements of ferromagnetic Fe$_{0.25}$TaS$_{2}$.  We observed that MO signal strongly depends on the photon energy, temperature, and magnetic field.  The Kerr angle exhibited pronounced peaks at $\sim$775~nm (1.6~eV) and $\sim$515~nm (2.4~eV), which we explain in terms of spin-dependent interband optical transitions involving electronic states in the vicinity of the Fermi energy.  At each fixed wavelength, the magnetic-field dependence of the Kerr angle showed a clear hysteresis loop, but, strikingly, the shape of the hysteresis loop was strongly wavelength dependent.  Based on a model taking into account the contributions from domain walls, we derived a mathematical expression that successfully fits all the observed hysteresis loops.

\section{Sample and Experimental Methods}\label{sec:Exp}

Single crystals of Fe$_{0.25}$TaS$_{2}$ were prepared by iodine vapor transport reaction in a closed silica tube, as described in Ref.~\onlinecite{MorosanPRB2007Fe}.  The sample exhibited strong magnetic anisotropy  with easy axis parallel to the crystallographic $c$ axis and had a ferromagnetic transition temperature $T_c$ of $\sim$160~K.

%
MOKE measurements were performed in the polar geometry in which the light beam was nearly normal incident on the sample surface.  Details of the experimental setup are described in Ref.~\onlinecite{CJ-GaMnAs-11-PRB}.  White light from a 100~W Xe lamp was first focused into a monochromator. Light with a selected wavelength was then polarized with a Glan-Thompson polarizer and impinged on the sample.  The reflected beam passed though a photoelastic modulator (PEM) and an analyzer, and then its intensity was detected with a Si photodiode.  The current signal from the photodiode was amplified and converted into a voltage and fed into two lock-in amplifiers.  The two lock-in amplifiers were used to demodulate the signal.  The first lock-in amplifier was referenced to the chopper frequency to provide a measurement of the average light intensity at each wavelength. The second lock-in amplifier was referenced to the second harmonic of the PEM frequency to record the fast oscillating signal at 100~kHz.  The Kerr rotation angle ($\theta_K$) was derived from the ratio of the two.  The sample was kept in a helium-flow cryostat, allowing us to vary the temperature ($T$) from 10~K to 300~K.  An external magnetic field was applied perpendicular to the sample surface and swept within the range between $-$2000~Oe and +2000~Oe.  The Faraday rotation induced by the cryostat window was subtracted.  Any polarization anisotropy caused by components in the setup was carefully calibrated and subtracted to get the accurate Kerr rotation at each wavelength.

\section{Experimental Results}\label{sec:Results}

\begin{figure}
\includegraphics [scale=0.65] {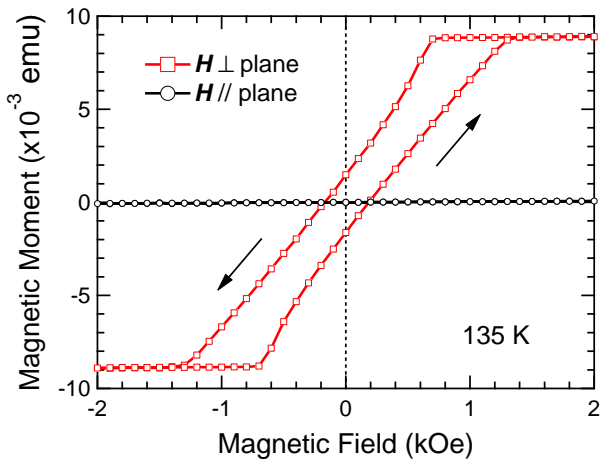}
\caption{(color online) Strongly anisotropic magnetization versus magnetic field ($H$) for the Fe$_{0.25}$TaS$_2$ sample at 135~K measured with $H \parallel c$-axis (red squares) and $H \parallel ab$-plane (black circles).  Magnetic moment is negligible in the $ab$ plane, indicating that the easy axis is parallel to the $c$ axis.} 
\label{SQUID}
\end{figure}

Magnetic moment versus magnetic field ($H$) curves at 135~K are shown in Fig.~\ref{SQUID}.  The moment is negligibly small when the field is applied parallel to the basal plane.  When $H$ is perpendicular to the sample plane, a strong magnetization with an oblique hysteresis loop is observed, magnetic switching occurring gradually over a range extended from $-1000$~Oe to $+1000$~Oe.  In the down-sweep direction, magnetization switching starts near 700~Oe and completes near $-1300$~Oe, with a coercive field of $\sim200$~Oe.  The remanent magnetization (at zero field) $M_r$ was about 17\% of the saturation magnetization $M_s$.  This switching behavior indicates that at this temperature (135~K) opposite domains nucleate before the field $H$ changes its sign.  Complete magnetization reversal consists of the formation and expansion of these opposite domains.  This is consistent with the MO imaging results of Fe$_{0.25}$TaS$_{2}$.\cite{VannettePRB2009}

%
\begin{figure}
\includegraphics [scale=0.6] {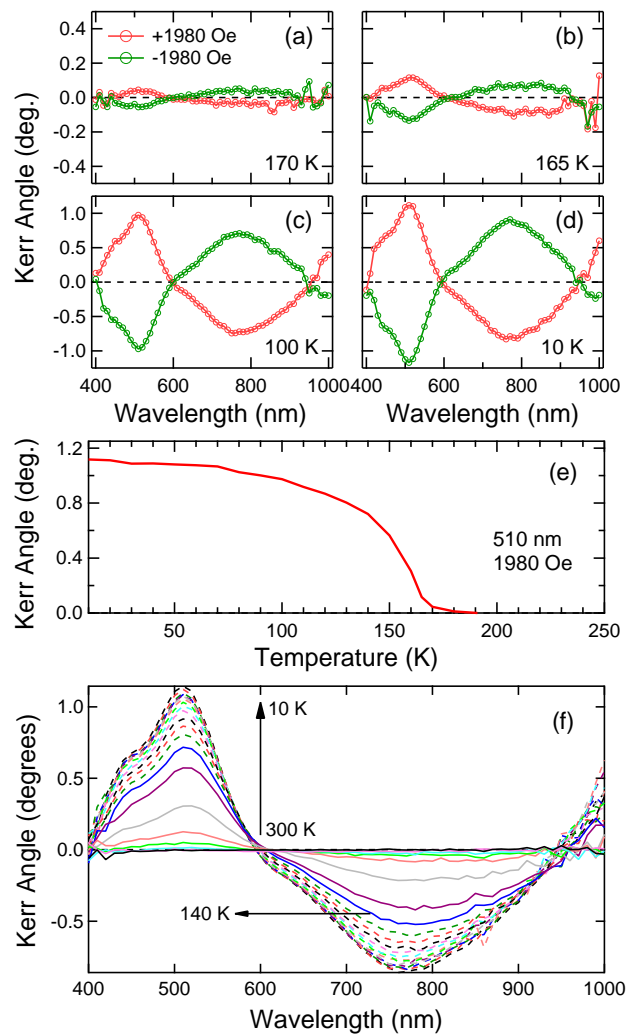}
\caption{(color online) MOKE spectra at (a) 170~K, (b) 165~K, (c) 100~K, and (d) 10~K in a magnetic field of $\pm1980$~Oe.  Temperature dependence of field-cooled (e) MOKE signal at 510~nm and (f) MOKE spectra in a field of $1980$~Oe from 300~K to 10~K.}\label{FC}
\end{figure}
%
Representative MOKE spectra measured at 170~K, 165~K, 100~K, and 10~K are shown in Fig.~\ref{FC}.  The sample was kept in a constant magnetic field ($+1980$~Oe or $-1980$~Oe) when the temperature was lowered from room temperature to 10~K.  The spectra obtained in the positive and negative fields are symmetric about zero, as expected.  The signal is absent at 300~K, starts appearing at 165~K, close to $T_c$, and increases with further decreasing temperature.  The MOKE signal amplitude is largest near 510~nm and 770~nm and zero around 400~nm, 600~nm, and 950~nm.  These largest and zero wavelengths do not change as a function of temperature.  Figure \ref{FC}(e) shows the MOKE signal at 510~nm taken while the sample was cooled in the presence of a field of $1980$~Oe.  Figure \ref{FC}(f) shows detailed temperature-dependent MOKE spectra in a field of $1980$~Oe from $T$ = 300~K to 10~K.

\begin{figure}
\includegraphics [scale=0.65] {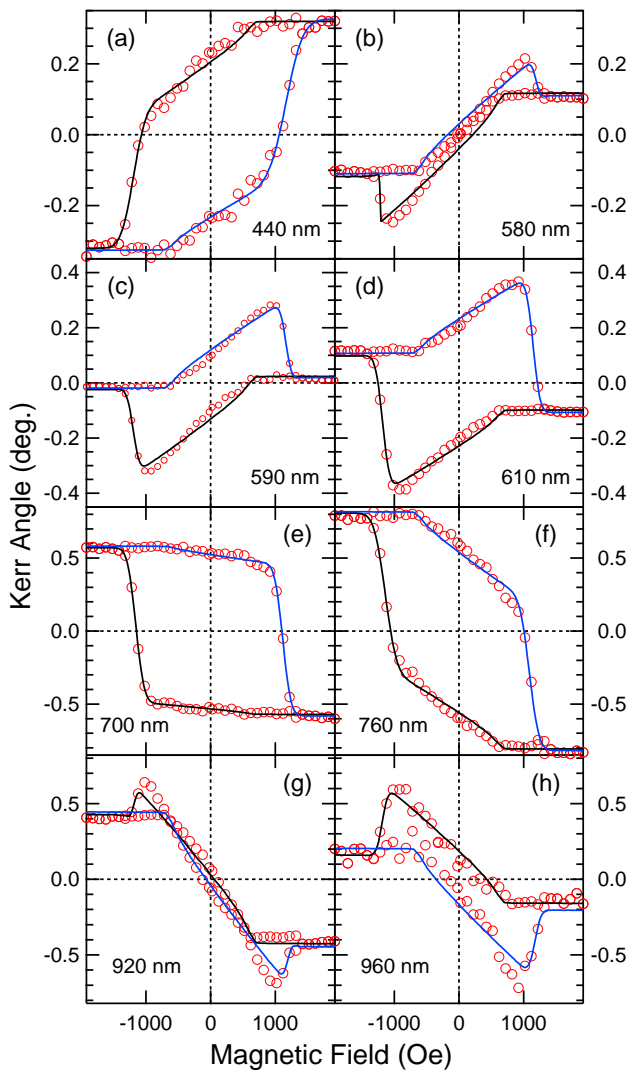}
\caption{(color online) MOKE hysteresis loops measured at various wavelengths at a constant temperature of 135~K.  The shape of the hysteresis loop changes drastically with wavelength.  Solids curves are fits to the data, using Eq.~\ref{eq:eqMOKE}, as discussed in detail in the text.}\label{loops}
\end{figure}

The Kerr rotation signal was observed to vary in a highly complex and unusual manner when the magnetic field was swept, and the magnetic field dependence was sensitively wavelength dependent.  Figures~\ref{loops}(a)-(h) display representative Kerr hysteresis loops for Fe$_{0.25}$TaS$_{2}$ at selected wavelengths measured at a constant temperature of 135~K.  None of these hysteresis loops resemble the magnetization curve (Fig.~\ref{SQUID}).  In addition, none of them is similar to each other.  Even the effective ``coercive" field is seen to vary with the wavelength.  Furthermore, in some hysteresis loops, that at 590~nm, for example, the Kerr signal vanishes at the highest fields ($\pm$1980~Oe), but it exhibits a peak at intermediate fields ($\pm1000$~Oe).

\begin{figure}[h]
\includegraphics [scale=0.48] {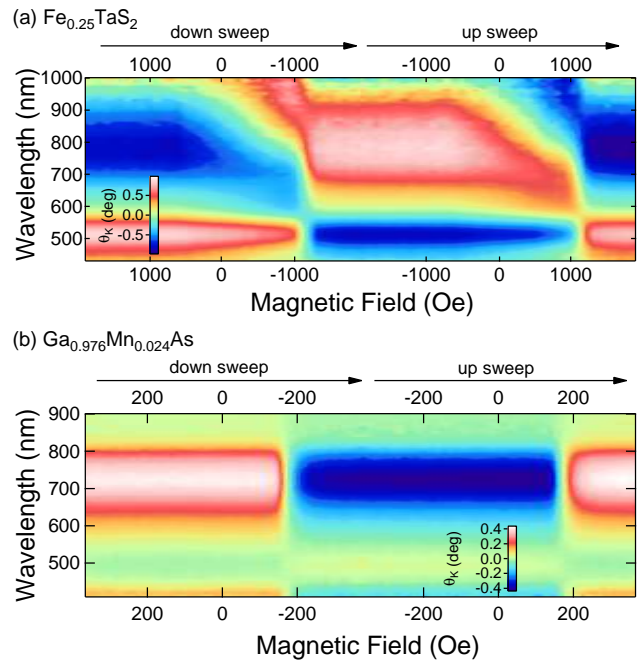}
\caption{(color online) (b) Contour plot of the measured Kerr angle as a function of magnetic field and wavelength for Fe$_{1/4}$TaS$_{2}$ at 135~K.  Magnetic field was swept from $+1920$~Oe to $-1920$~Oe, and back to $+1920$~Oe, completing a full cycle.  (b) Data for a reference Ga$_{0.976}$Mn$_{0.024}$As sample showing standard hysteresis curves with no anomalies.}
\label{contours}
\end{figure}

To further highlight these unusual wavelength-dependent hysteresis curves, Fig.~\ref{contours}(a) displays a contour map of the Kerr rotation angle as a simultaneous function of magnetic field and wavelength, obtained at 135~K.  Here, as we go from the left to right, the magnetic field changes from $+1920$~Oe (left end) to $-1920$~Oe, and then back to $+1920$~Oe (right end).  The Kerr signal shows a complex pattern as a function of magnetic field, and the details sensitively change from wavelength to wavelength.   As a comparison, Fig.~\ref{contours}(b) presents MOKE data for ferromagnetic Ga$_{0.976}$Mn$_{0.024}$As in the same scheme as in Fig.~\ref{contours}(a).  It is clear that the hysteresis shape is constant in the case of Ga$_{0.976}$Mn$_{0.024}$As, showing a wavelength-independent coercive field ($\sim$180~Oe), even though the signal size changes with the wavelength.

\section{Discussion}\label{sec:Discussion}

\subsection{Photon-energy-dependent Kerr angle}

Within the framework of band structure theory, the photon energy dependence of magneto-optical Kerr signal is related to the difference of joint density of states (JDOS) between the spin-up and spin-down bands.  Simulations of MOKE spectra require the knowledge of the band structure of the material.  There exists no report on calculating the band structure of Fe$_{0.25}$TaS$_2$, to our knowledge, while band structure calculations have been reported for similar compouns Fe$_{1/3}$TaS$_2$, Mn$_{1/3}$TaS$_2$, and Mn$_{1/4}$TaS$_2$.\cite{DijkstraJPCM1989,WijngaardJP1988,MotizukiJMMM1992}  The general features that are true for this group of materials are: (1) large spin splitting of transition metal (Fe or Mn) 3$d$ bands; (2) hybridization of transition metal 3$d$ and Ta 5$d_{z^{2}}$ states; and (3) large DOS of transition metal 3$d$ and Ta 5$d_{z^{2}}$ at the Fermi energy.  Thus, the transitions that contribute to the Kerr signals are those from the Fe 3$d$ and Ta 5$d_{z^{2}}$ near the Fermi energy to higher lying Ta 5$d$ states.

We adopted the DOS results from Ref.~\onlinecite{DijkstraJPCM1989} to get an estimate of MOKE spectra of Fe$_{0.25}$TaS$_2$ and compared that with our experimental data for fully magnetically-polarized samples.  The spin-split DOS of Fe$_{1/3}$TaS$_2$ (not shown) can be decomposed into those for Fe, Ta(1), Ta(2), and S atoms.  Here, Ta(1) and Ta(2) represent the two types of Ta atoms in the crystal lattice: Ta(1) atoms have no direct Fe neighbors, whereas each Ta(2) atom has one Fe neighbor.  We make the following assumptions: i) transitions occur within each kind of atoms, i.e., Fe $\rightarrow$ Fe, Ta(1) $\rightarrow$ Ta(1), and $\rightarrow$; ii) transition rate T equals JDOS; and iii) Kerr signal is proportional to the difference of spin-up and spin-down transition rate. Assumption i) neglects the hybridization between the bands of different atoms.  Assumption ii) neglects the different matrix elements for different kinds of atoms.  Assumption iii) neglects the wavelength dependence of the proportionality coefficient.

\begin{figure}
\includegraphics [scale=0.55] {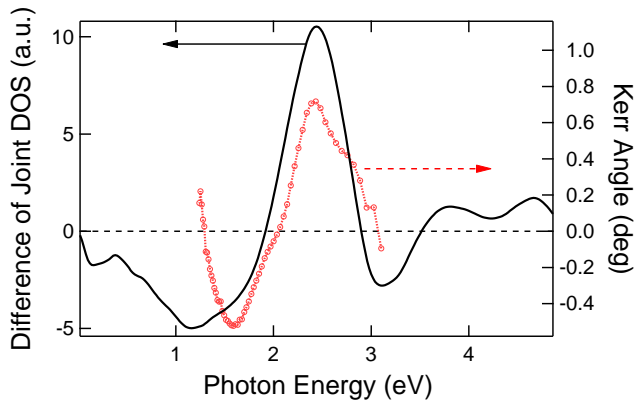}
\caption{(color online) The difference in JDOS between the spin-up and spin-down bands calculated for Fe$_{0.25}$TaS$_2$ (left axis), together with the experimentally measured MOKE spectrum at 140 K in an external magnetic field of 1980 Oe (right axis).}
\label{JDOS}
\end{figure}

Using these results and the above assumptions, the JDOS was calculated through
\begin{equation}
{\rm JDOS}(E) = \int^{+5~{\rm eV}}_{-5~{\rm eV}} {\rm DOS}(E_1) \cdot {\rm DOS}(E_1 + E)dE_1.
\end{equation}
The calculated JDOS is plotted in Fig.~\ref{JDOS}, together with the experimental MOKE spectrum at 140~K.  There is qualitative agreement between them, and the essential spectral features in the experimental spectrum are correctly captured by the calculation.  In particular, there is a peak at 2.4~eV in both cases.

\subsection{Anomalous hysteresis loops}

It is usually true that the MOKE angle measured in the polar Kerr geometry is proportional to the magnetization ($M$) perpendicular to the sample surface.\cite{Shinagawa}  However, this general rule is obviously not held in our data, comparing the magnetization curve (Fig.~\ref{SQUID}) and MOKE hysteresis curves (Fig.~\ref{loops}).

\begin{figure}
\includegraphics [scale=0.65] {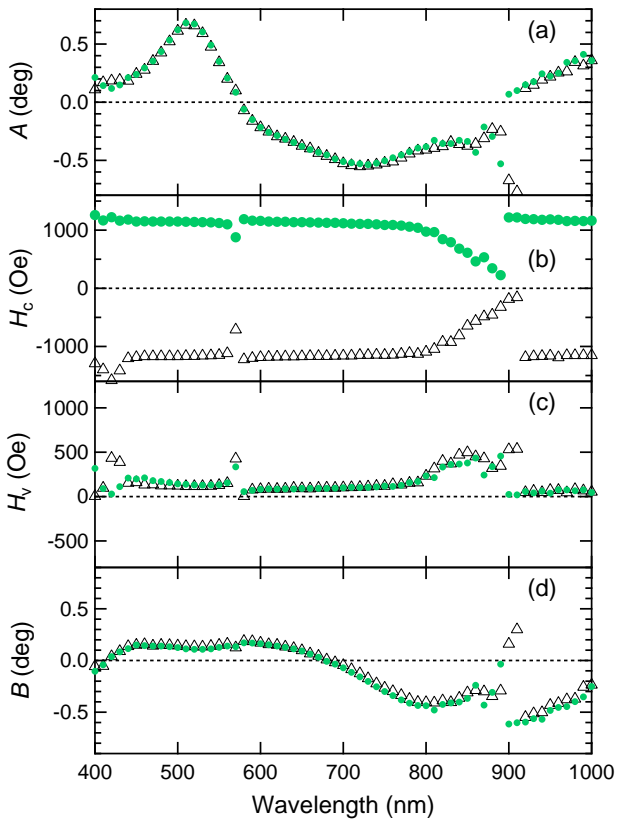}
\caption{(color online) Spectral dependence of the fitting parameters used in Eq.~(\ref{eq:eqMOKE}) to fit the MOKE hysteresis loops in Fig.~\ref{loops}. Green circles and black triangles represent the down-sweep and up-sweep, respectively.}\label{fitPara}
\end{figure}

To explain the observed abnormal Kerr hysteresis loops, we propose a contribution to the Kerr signal that is additional to the part proportional to $M$.  We use the following fitting function for the MOKE signal:
\begin{equation}\label{eq:eqMOKE}
\textrm{MOKE}(H) = A \cdot \textrm{erf} \frac{H-H_{c}}{\sqrt{2}H_{v}} + B \cdot {M(H) \over M_s},
\end{equation}
where $A$, $H_{c}$, $H_v$, and $B$ are fitting parameters, while $M(H)$ is the magnetization and $M_s$ is the saturation magnetization value; $M(H)/M_s$ versus $H$ can be obtained from the data in Fig.~\ref{SQUID}.  Applying Eq.~(\ref{eq:eqMOKE}) to fit all the MOKE hysteresis loops, we obtained approximately wavelength-independent fields $H_c\approx1.2\pm0.1$~kOe and $H_v\approx0.10\pm0.03$~kOe.  The fitting curves reproduce the observed hysteresis loops for all the wavelengths, as shown as solid lines in Fig.~\ref{loops}.  Figure~\ref{fitPara} shows the dependence of the fitting parameters on the wavelength.

\subsection{Microscopic model}

In order to get a better understanding of the data, we propose the following model. During a field scan, spin-up domains and spin-down domains coexist. Let us assume that $V_+$ out of the total sample volume $V_0$ is occupied by spin-up domains, and define a partition number $f = V_+/V_0$. Then, the spin-down domain volume is $V_- = V_0 - V_+$ with partition number $(1 - f )$. The total magnetization can be expressed as 
\begin{equation}
M = \mu \times (V_+ - V_-) = \mu V_0 (2f - 1) = M_s(2f - 1),
\end{equation}
where $\mu$ is the magnetic moment per unit volume and $M_s$ is the saturation magnetization.

Since $M$ is a function of $H$, $f$ is also a function of $H$:
\begin{equation}
f(H) = {1 \over 2} \left ( {M(H) \over M_s} + 1 \right ).
\end{equation}
Let $\theta_K$ be the microscopic Kerr angle in a single domain.  If domains were large and domain walls constituted only a small fraction of the sample, then one could neglect the dependence of $\theta_K$ on the size of the domain.  However, due to the special dendritic structure of the domains in the present material,\cite{VannettePRB2009} domain walls occupy a significant fraction of the total sample size, and their effect cannot be neglected.  Let us introduce $\theta_K(V_{\pm}/V_0)$,  which corresponds to averaged Kerr angles  for positive and negative domains, taking into account effects of domain walls.  Since the relative volume fraction of domain walls depends on $f$, the averaged Kerr angles will also depend on it.  The measured Kerr angle $\Theta_K$ is the result of averaging over both types of domains:
\begin{equation}
\Theta_K = f \theta_K(V_+/V_0)+(1-f)\theta_K(V_-/V_0).
\label{Theta_K}
\end{equation}

Because $\theta_K$ is proportional to the difference of absorption coefficients between right and left circularly polarized light, i.e., $(\alpha_+ - \alpha_-)$, it is determined by the difference of joint density of states of spin-up and spin-down bands. Consider the three situations: (1) $1\gg f >0$, (2) $f$ =1, and (3) $0<f <1$.  In the first case, a small spin-up(+) domain is nucleated inside a single spin-down(-) domain.  Due to the finite size of the positive domain and the nearby domain walls, the spin-split band structure can be shifted compared to the bulk sample consisting of a single positive domain.  Therefore, the absorption coefficients have {\em reduced} values $\alpha_{\pm}^r$ , resulting in a reduced Kerr angle $\theta_K^r$.  In the second case, a single domain of spin-up is formed. The band structure is the same as the bulk sample. Thus, $\alpha_{\pm}$ and $\theta_K$ have the normal values $\alpha_{\pm}^0$ and $\theta_K^0$.  In the general case (3), $\alpha_{\pm}$ and $\theta_K$ are between the reduced and normal values.  Let us assume that the change of $\theta_K$ from the reduced value $\theta_K^r$ to the normal value $\theta_K^0$ can be described by an error function
\begin{equation}
{\rm erf}(\theta) =  {\rm erf}{{H - H_{c\theta}} \over \sqrt 2 H_{v \theta}}.
\end{equation}

There is no microscopic reason to justify this particular type of dependence on the magnetic field, but we expect that any smooth function that can interpolate between $\theta_K^r$ and $\theta_K^0$ will capture the effect of domain walls.  Defining
\begin{eqnarray}
\theta^S_K &=& {1 \over 2} (\theta^0_{K} + \theta^r_{K}), \\
\theta^A_K &=& {1 \over 2} (\theta^0_{K} - \theta^r_{K}),
\end{eqnarray}
we can express
\begin{eqnarray}
\theta_K (V_+/V_0) &=&  \theta^S_K + \theta^A_K \times {\rm erf}(\theta), \\
\theta_K (V_-/V_0) &=&  -\theta^S_K + \theta^A_K \times {\rm erf}(\theta).
\end{eqnarray}
Substituting the above in Eq.~(\ref{Theta_K}), we have
\begin{eqnarray}
\Theta_K &=&	f (\theta^S_K + \theta^A_K {\rm erf}(\theta)) + (1 - f )(-\theta^S_K + \theta^A_K {\rm erf}(\theta)) \nonumber \\
&=&	\theta^A_K {\rm erf}(\theta) + (2f - 1)\theta^S_K \nonumber \\
&=&	\theta^A_K {\rm erf}(\theta) + \theta^S_K	{M(H) \over M_s}.
\label{adilet-eq}
\end{eqnarray}
We immediately see that Eq.~(\ref{adilet-eq}) agrees with our fitting equation [Eq.~(\ref{eq:eqMOKE})].  The consistency of our interpretation of  the origin of fitting terms in Eq.~(\ref{eq:eqMOKE}) also implies that $H_c$ and $H_v$ are roughly independent of the wavelength, which is demonstrated by the extracted fitting parameters in Fig.~\ref{fitPara}.

\section{Conclusions}\label{sec:Conclusion}

We have used magneto-optical Kerr spectroscopy to investigate Fe$_{0.25}$TaS$_2$ with out-of-plane anisotropy as a function of temperature (10~K to room temperature), magnetic field ($\pm$1980~Oe), and photon energy ($\sim$1.4-3~eV).  We attributed the optical Kerr signal to the interband transitions and explained the spectra with the difference of the joint densities of states of spin-up and spin-down bands.  This successfully captured the spectral peak position and qualitatively reproduced the experimental results.  At a fixed wavelength, the magnetic-field dependence of the Kerr angle showed a clear hysteresis loop, but its shape sensitively changed with the wavelength.  We proposed a simple mathematical expression for fitting all the hysteresis loops and provided a physical description based on domain wall physics.  

\begin{acknowledgments}
This work was supported by the NSF through Award No.~OISE-0530220 and the Alfred P.~Sloan Foundation.  We thank Hiro Munekata for providing the reference GaMnAs sample used for Fig.~\ref{contours} and Ruslan Prozorov for useful discussions.
\end{acknowledgments}


\end{document}